# Wind Turbine Generated Power Enhancement by Plasma Actuator

Javad Omidi [1]
Research Assistant, Aerospace Engineering Department,
Sharif University of Technology, Tehran, Iran
jomidi@alum.sharif.ir

Karim Mazaheri [2]
Professor, Aerospace Engineering Department,
Sharif University of Technology, Tehran, Iran
mazaheri@sharif.ir

*Abstract*— Here we have investigated the usage of a Dielectric Barrier Discharge (DBD) plasma actuator to improve the aerodynamic performance of an offshore 6MW wind turbine. By controlling the aerodynamic load at different blade sections, the overall generated power and the blade loading is improved. The blade is divided to seven segments and the analysis is applied to each segment. To study effects of configuration and location, actuators were installed in a single or tandem configuration, spread between 46% and 56% of the span length, at different chord wise locations. The improved phenomenological model developed by authors were used to simulate the interaction of the electrostatic field, the ionized particles and the fluid flow. Using the aerodynamic coefficients of each section predicted by numerical simulation, the generated power was estimated. We investigated how the location and configuration of the single and tandem actuators improve the generated power up to 95kW. Also, we studied possibility of using collective pitch angles. It showed that with a collective pitch angle of 10 degrees, one may harvest 10.03% or 11.42% more energy at wind speeds of 6 and 10 m/s respectively. Among seven cases studied, actuators closer to the leading edge performed better than others.

## I. BACKGROUND

Energy harvesting using reusable energies, as the most reliable source for provision of the energy in the future, has absorbed most of the current research interest in energy production and harvesting [1, 2]. Techniques studied in these researches include many different methods and each research is devoted to improve the efficiency of energy harvesting methods. For obvious reasons, wind energy is one of the most interesting and for a long time it has been used by humans to produce mechanical energy. The wind industry suffers from the production cost, maintenance and repair expenses and limited life of turbine structures. Another challenge is to find wind farms for turbine installations for which a continuous wind is guaranteed. To address this challenge, here we use Dielectric Barrier Discharge (DBD) actuators to control flow around turbine blades, to make it possible to produce the same power for a lower wind speed which is equivalent to harvest a higher power in a similar wind speed.

During the last decade, the control of the flow around aerodynamic bodies has significantly improved. One of the main improvements is devoted to flow control around wind turbine blades. Many researches are going on to optimize the aerodynamic load over the wind turbine blade [3, 4, and 5]. Since experimental investigations in this field are very time consuming and expensive, using new algorithms and high speed computing facilities motivates usage of computational techniques to simulate the flow field around turbine blades to achieve a better control of the flow field [6].

Because of their simplicity and effectiveness, usage of plasma actuators and specially DBD actuators has received special attention in recent years. Its advantages include low weight, easy installation, using no moving parts, no need to pneumatic, hydraulic systems and its low power consumption [7]. Its low cost and relatively easy simulation make it more suitable for engineering applications [8].

Many people have investigated application of DBD actuators experimentally. Roth et al [9] have studied usage of plasma actuators to control the boundary layer. Corke et al [10] have studied using these actuators to stimulate instability of the boundary layer over a pointed cone at Mach 3.5 and also Corke and Post [11, 12, and 13] have investigated leading edge separation control for NACA66$_3$-018 and NACA0015 airfoils to avoid static and dynamic stalls. Jacob et al [14] considered their usage for laminar and turbulent flow control over a flat plate and also for low pressure gas turbine blades. Orlove et al [15] studied leading edge separation control for NACA0021 airfoil at different angles of attack after stall. He et al [16] studied using Hump model to analyze the effect of plasma actuators on flow control. Jolibois et al [17] optimized the location of an actuator on a NACA0015 airfoil for post-stall angles of attack. Little et al [18] considered usage of a plasma actuator on a NASA EET airfoil flap. Finally, Thomas et al [19] used a plasma actuator to optimize the lift coefficient. These are only a few among reported experimental works in this field, which show an extensive interest to understand how plasma actuators affect the flow field and to utilize it in engineering applications.

Another line of contemporary research in this field is devoted to the improvement of the aerodynamic properties of wind turbine blades using plasma actuators. Nelson et al [20] have used numerical and experimental methods to design an intelligent blade, using plasma actuators which control the local flow separation to preserve lift coefficient after the stall point. Versailles et al [21] investigated experimentally and numerically the potential usage of plasma actuators to improve blade lift coefficients and to control the output turbine power in high-wind situations. Walker et al [22] studied their usage for boundary layer separation control around NACA0024 airfoil to optimize performance of wind turbine blades at high angles of attack.

To optimize performance of wind turbine blades, many different control algorithms and strategies are implemented to improve the performance for different environmental conditions. Grennblatt et al [23, 24] investigated dynamic flow separation control and performance improvement for vertical axis wind turbine using plasma actuators at their leading edges. In another experimental study, Grennblatt et al [25] used a plasma actuator to control the dynamic stall on a double bladed H-rotor vertical axis wind turbine with 7 m/s air-speed. They measured power coefficients for different speed ratio and used a feed-forward control to have a constant load and achieved a net 10% power increase. Baleriola et al [26] used an open-loop circulation control strategy to control the aerodynamic load actuations on a wind turbine blade using plasma actuators.

Cooney et al [27] used an active flow control strategy and three types of plasma actuators to reduce the levelized cost of the wind energy. They designed actuators to be mounted on existing blades or to be embedded in new blades. Batlle et al [28] developed a multi-objective optimization design methodology for design of a DBD actuated airfoil with active flow control. Two cost functions, one for wind energy suitability to find the optimum airfoil geometry and one for actuator control suitability to optimize the actuator parameters, were used and was experimentally validated to find a suitable actuated blade. Hisashi Matsuda et al [29] used CFD analysis to study mounting a DBD actuator on a 1.75 MW wind turbine blade and achieved a higher blade rotational speed at the same wind-speed and an average 4.9% generated power increase.

Sandrine Auburn et al [4] reviewed active flow control strategies for wind turbines. Haitian Zhu et al [5] reviewed jet flow control techniques including the blowing, synthetic and plasma actuators, for vertical axis wind turbines and concluded that their energy and matter consumption shall be reduced to make them economically valuable. Li Guoqiang et al [30] experimentally studied usage of DBD actuators for large wind turbines to control their blade airfoil dynamic stall. They used synchronous pressure measurement to trigger PIV (particle image velocimetry) tracking. They were able to increase the aerodynamic efficiency and to reduce the hysteresis loop region generated by the cyclic pitch variation.

Ebrahimi et al [31] have conducted a numerical simulation to analyze performance of a 5 MW NREL wind turbine blade. They have used a very simple electrostatic model without considering effects of geometry of the actuator and operation parameters (e.g. voltage, frequency, etc.) and without validation for curved surfaces, since that model is not valid for curved surfaces. They have considered installation of the actuator close to the hub which is practically insignificant for generated moment or power harvesting. To find the most appropriate section for studying the actuator performance is one important aspect, which will be noticed here.

To use computational schemes in optimization of wind turbine blades, requires usage of phenomenological models, since direct simulation of the governing equations, even for a single design, is still far from our capabilities. Omidi [3, 6] used existing simulation models to simulate interaction of a DBD actuator with the fluid flow around a flat plate and also curved surfaces like two dimensional cylinders and thin and thick airfoils. He optimized the location of the DBD actuators for best performance. He also used the electrostatic model of Suzen and Huang [32, 33] and the improvement proposed by Bouchmal [34] to control the von Karman vortex shedding on a two dimensional cylinder at a Reynolds number of 20000. Mounting two actuators on the upper and lower surfaces of the cylinder decreased the lift fluctuations up to 55% and also decreased the average drag coefficient up to 90%. Authors [3] used published experimental results to estimate Debye length (developed plasma) over a dielectric surface, including effects of the actuator geometry, voltage and frequency. They also used a combination of boundary conditions used in different works and introduced a new phenomenological model. This was applied to analyze effects of the voltage and frequency on an actuator performance and to study effect of actuator location on the performance of a wind turbine section for a wide range of angle of attacks (AOAs). Finally, they improved the phenomenological model [6] and made it completely independent of experimental results for calibration of the model parameters.

Here we have investigated application of DBD actuators to improve performance of a huge 6MW three dimensional wind turbine blade. We used validated models introduced by ourselves [3, 6] to understand the physics of flow interactions and to use them to increase the generated power in similar wind-speeds or to decrease the turbine rotational speed for same harvested power. The main advantage was to increase the generated power when the wind speed is too low. We used detailed analysis for one section of the blade and investigated seven different single or tandem actuator configurations to improve the performance of this section. Further we used a wind turbine blade design software and some simplifying assumptions to estimate the power increment by using DBD plasma actuators on all sections of a full scale 6 MW wind turbine blade and studied the possibility of using a collective pitch angle to improve energy harvesting.

## II. IMPLEMENTATION

### A. Problem Description

Here we investigated how to improve the aerodynamic performance of an offshore DOWEC 6MW wind turbine blade [35], using a DBD actuator. The turbine rotor has three blades with a diameter of 129 m and its design wind speed is 11.4 m/s. Table 1 gives details of specifications of the turbine. A pitch control system regulates the output power for different wind speeds. Each blade is made up of seven different airfoils, as shown in Table 2. Table 2 also gives the Reynolds number for design wind speed at different blade sections. Here we have investigated installation of one or two DBD plasma actuators between two stations, located at 46% and 56% of the blade span.

**Table 1- DOWEC 6MW wind turbine specifications [35]**

| Rating Power | 6 MW |
|---|---|
| Rotor Diameter | 129 m |
| Design Tip Speed | 80 m/s |
| Cut in/out Speed | 3 m/s, 25 m/s |
| Rated Wind Speed | 12.1 m/s |
| Hub Height | 90 m |
| Rotor, Hub Diameter | 126m, 3m |
| Rotor Configuration | Upwind, 3 Blades |
| Control | Variable Speed, Collective Pitch |
| Drivetrain | High Speed, Multiple-Stage Gearbox |

**Table 2- Seven different airfoil sections and their Reynolds number for the design wind speed**

| r (m) | Airfoil | Range of Airfoil (m) | Re × $10^{-6}$ |
|---|---|---|---|
| 10.5 | DU40 | 10.04 – 13.80 | 4.436 |
| 18 | DU35 | 13.80 – 22.00 | 6.501 |
| 25.5 | DU30 | 22.00 – 26.10 | 7.898 |
| 33 | DU25 | 26.10 – 34.30 | 8.875 |
| 40.5 | DU21 | 34.30 – 42.50 | 9.419 |
| 48 | NACA64 | 42.50 – 52.75 | 9.505 |
| 55.5 | NACA64 | 52.75 – 63.00 | 9.221 |

The flow complexities around the airfoil close to the hub motivates a higher priority for flow control, but, with a short arm, contribution of this part of the blade in the generated power is not significant. Therefore a mid-section is considered in this study. For location of a tentative single DBD actuator configuration, three chord-wise locations (40%, 50%, and 60% of the chord) from the leading edge was used. Similarly, for tandem configurations, four different locations, i.e. 30-40, 40-50, 50-60, and 60-70% of the chord were used. Based on our previous experience, the voltage applied to the actuator is 12kV and the frequency was set to 5.5 kHz. We have used computational aerodynamic analysis to find the aerodynamic performance of each section, with or without the DBD actuator and then we would compare the original and the improved generated power.

### III. GOVERNING EQUATIONS

*A. Hydrodynamics Equations*

Here we use two-dimensional RANS equations to simulate the incompressible fluid flow [6]. Since most of the energy given to the fluid by the actuator accelerates the fluid particles and has little contribution in warming of the flow field [8], the energy equations are not considered and only the continuity and the momentum equations are solved here:

$$\nabla . u = 0 \quad (1)$$

$$(u.\nabla)u = -\frac{1}{\rho}\nabla p + \vartheta\nabla^2 u + f_{b_i} \quad (2)$$

Here, $p$ is the static pressure in Pa, $\vartheta$ is the fluid kinematic viscosity in m²/s, $u_j$ is the fluid velocity in m/s, $\rho$ is the dencity in kg/m³ and $\overrightarrow{f_b}$ is the body force per unit volume, generated by the actuator, in N/m³. The body force is added as a source term to the momentum equation, equation (2).

*B. Simulation of the Fluid and Actuator ElectroHydroDynamic Interactions*

Many electro-hydrodynamic models are suggested for modeling plasma actuators, which simulates the generated plasma and the electrical potential field and add the interaction force as a source term to the momentum equation. The model presented by Roth [36], The electrostatic model of Suzen, Huang et al [32, 33], the lumped element circuit model of Orlove et al [37], the linearized force model of Jayaramann and Shy [38], the potential flow model of Hall [39] and the complete model of solving the Maxwell's equations without neglecting the magnetic terms [40] are probably the most important models proposed during the last two decades for modeling the body force of the plasma actuators. The authors have presented a detailed review in [3, 6] and have introduced an improved model based on the model of Suzen and Huang. This model, which is used here, is an appropriate engineering model to predict the electro-hydrodynamic interactions with a quite good accuracy.

When a high (1-18 kV) voltage is applied to two electrodes of a DBD actuator, a continuous rapid ionization is happened with a time scale of nanoseconds. This ionized fluid is continuously generated and is washed down by the moving fluid [34]. The charged fluid particles inside the electrical field are forced by the electrical field and exert an average body force on the fluid flow. The flow timescale is about milliseconds; therefore, the hydrodynamic flow field only senses an average body force and we do not need to consider the nanosecond details of this interaction. Neglecting the magnetic forces, the Lorentz body force [32, 33] is:

$$\overrightarrow{f_{b_j}} = \rho_c \overrightarrow{E_j} \quad (3)$$

in which $\rho_c$ is the plasma dencity in terms of C/m³ and $\vec{E}$ is the electrical field vector. The model of Suzen and Huang is one of the best physical models and assumes enough time for continuous generation of plasma as a quasi-steady process. The gas particles are weakly ionized in the process of plasma generation [41], therefore Suzen and Huang [32] assumed that the electrical potential is made up by linear superposition of two potentials: the potential of the electrical field and the potential of the charged particles. However, the Debye length is small enough and the plasma layer is very thin, therefore they assumed that the plasma density is mostly affected by the potential of the charged particles over the wall, instead of the electrical field. They used the superposition principle to find equations for the electrical field and the charge density distribution based on the Maxwell's equations:

$$\nabla . E = \nabla .(\varepsilon_r \nabla \phi) = 0 \quad (4)$$

$$\nabla .(\varepsilon_r \nabla \rho_c) = \frac{\rho_c}{\lambda_d^2} \quad (5)$$

in which $\phi$ is the electrical potential, $\lambda_d$ is the Debye length and $\varepsilon_r$ is the relative electrical diffusivity. The electrical potential equation (4) is solved both in the Dielectric part and also in the flow simulation part. The boundary condition applied to the outer boundary is $\partial\phi/\partial n_i = 0$, over the external electrode surface is $\phi = \phi(t)$ and over the internal electrode surface is $\phi = 0$. $n_i$ is the unit normal vector and $\phi(t) = \phi^{max} f(t)$ is the applied alternative voltage. The charge density equation (5) is only solved in the fluid simulation part. The boundary condition for the external boundary is $\rho_c = 0$ and on the internal electrode surface is $\rho_c = \rho_c(x) f(t)$ and over other surfaces of the dielectric and the external electrode is $\partial\rho_c/\partial n_i = 0$. More details regarding these equations may be found in [2, 3, 32, and 33].

### C. Numerical Methods

ANSYS Fluent 15.0 was used to solve both the electrostatic model and the fluid flow equations. Since the flow is a hybrid laminar and turbulent flow, SST transition turbulence model [42] was used. The second order SIMPLE algorithm was used for all variables to couple velocity and pressure. The solution procedure was continued to reach 13 order of convergence for the continuity equation and 7 order of convergence for all other residuals, in addition to convergence of the aerodynamic coefficients. Also the difference of the inflow and the outflow mass flux was considered. Details of the numerical method are given in [3, 6].

### D. Solution Domain and Grid

The flow over the leading edge was assumed to be laminar and attached. Since the wind turbine blades are supposed to work at different wind speeds, they shall be productive at high angle of attack, or in other words, their stall point should be far enough. Therefore, the separation possibility in this flow is very high.

Meanwhile, since the Reynolds number is relatively high, the transition to turbulent flow on both airfoil surfaces, excited by the separation zone, are expected to happen and would have significant effect on our solution. The numerical solution model shall include capability of resolving separation and transition. An extensive part of both the lower and the upper airfoil surfaces (at least about 40% of the chord, for our solutions) is dominated by a laminar flow.

The grid quality is crucial for our numerical solution. The grid was generally selected based on the solution physical requirements, our algorithm, required solution resolution and our computational resources. The flow was assumed to be incompressible and the airfoil sections were relatively thick. As explained above, the grid should be suitable to resolve the boundary layer, the separation, the transition and the wake region. Also, since we install the DBD actuator on the suction surface and its location is variable and since the solution resolution near the DBD actuators is the most significant, a very fine mesh was required in the suction surface.

Finally, we needed to do many simulations and because of limitations in our computational power, we needed to produce a grid which is fine enough in critical regions of the flow field and at the same time as coarse as possible in other regions. For this reason, the flow field was decomposed to three regions with different requirements, each was separately discretized and all were integrated to find the most appropriate grid resolution.

The grid is C shaped and the joint boundaries are defined as interface. To help for a faster convergence in our incompressible flow field, the computational domain was stretched about 50 chords, but a coarse grid was selected far from the airfoil to reduce the computational cost. The front, the upper and the lower outer boundaries were assumed about 17 chords far from the airfoil surface and the back boundary is about 30 chords far from the trailing edge of the airfoil, to appropriately resolve the wake region. Grid resolution studies are reported in [3] and it was concluded that a grid with about 38000 cells has adequate accuracy for this analysis.

This zone including the airfoil is the most important region, with the finest mesh. Since we used the fairly more accurate turbulence model of SST Transition, the viscous sublayer and the buffer zone in the boundary layer should have been well resolved. It is recommended to have $y^+ < 1$ or even 0.5 and to have more than 10 cells in the boundary layer in this region [42]. Since y+ distribution over the airfoil in this grid was always below 0.1, it showed that the grid may resolve very accurately the viscous sublayer. Here we have solved the high angle of attack cases, therefore to resolve the separation zone, the number of grids normal to the airfoil surface should have been enough. Also, at least in part of the wake region, we needed enough grid resolution.

The geometrical dimensions of the DBD actuator follows. The dielectric thickness was 0.01 of the chord, the electrodes thickness was 0.0025 of the chord, the length of the external electrode was 0.07 of the chord and had the same external surface as the airfoil and the length of the internal electrode was 0.13 of the chord. This optimum configuration was recommended in [3].

## IV. RESULTS

The final performance index was the potential rotor generated power at each speed. We used PROPID-Ver53 [43] software to estimate the rotor generated power. This software is used for preliminary design of wind turbine blades. At best conditions, the clean rotor reaches its maximum rated power at about 12 m/s wind-speed. In fact this study may help us to increase the generated power for speeds well below this threshold velocity.

To have an estimate that how much the actuation may improve the harvested energy, we used PROPID53 software. First we assumed no change in the blade geometry and the pitch angle and estimated the energy increment at the rated velocity of 12.1 m/s. In this case, the angle of attack for each section is similar to the clean blade. Table 3 shows results for the increment in the generated power for all seven actuator configurations, in comparison with respect to the clean blade. The generated power for 30-40% tandem actuation is the most improved, which is about 95 kW. All tandem configurations, irrespective of their installation locations, perform better than all single actuators. One also observes that as the actuator is moved toward the trailing edge, its performance in the boundary layer control and power improvement is decreased. The minimum improvement is due to an actuator at 60% the chord length from the leading edge and this actuator has started to control the boundary layer when the flow has lost most of its momentum and the adverse pressure gradient has effectively de-energized the boundary layer.

**Table 3- the power increment estimated for different actuator configurations in comparison with the clean blade**

| Actuator Location | Type | Increment of Output Power (kW) |
|---|---|---|
| 40% | Single | 61 |
| 50% | Single | 45 |
| 60% | Single | 41 |
| 30% & 40%. | Tandem | 95 |
| 40% & 50% | Tandem | 84 |
| 50% & 60% | Tandem | 71 |
| 60% & 70% | Tandem | 63 |

The potential of increment in the total harvested energy, if the DBD actuators are installed in all sections is much higher. Four cases are studied. Table 1 shows details of all cases studied here. Effect of three collective pitch controls on total harvested energy for two different single (40%) and tandem (40-50%) actuators, each at both low and high air-speeds are estimated. For low speed estimates, we have computed results for air speed equal to 6 m/s and for high speed 10 m/s is used. Results are shown in Table 4. For each case, three different collective pitch angles of 2, 5, and 10 degrees are considered. Percent increments in the harvested energy are shown in Table 4. The best improvement potentials are in low speeds, since in high speeds, usually structural limits does not allow to increase the generated power, but in low speeds, the energy increment is easily harvested. Please note that the energy increment in low speed of 6 m/s for 10 degrees pitch angle is respectively 59.1KW (9.85%) and 62.1 KW (10.35%) for single and tandem configurations, respectively and these values are increased respectively to 341KW (10.03%) and 388.3 KW (11.42%) for air speed of 10 m/s. The above estimations show how application of DBD actuators on wind turbine rotors may improve their performance for low and high air speeds. In other words, using DBD actuators, one may harvest maximum allowable power for a wider range of air speeds.

**Table 4- increment in energy harvested in percent for different improvement scenarios respect to clean blade at zero pitch angle**

| Actuator Type | Wind Speed | Increment of Generated Power | | | |
|---|---|---|---|---|---|
| | | 0 Pitch Angle | 2º Pitch Angle | 5º Pitch Angle | 10º Pitch Angle |
| Single 40% | 6 m/s | 2.25 | 3.42 | 6.21 | 9.85 |
| | 10 m/s | 1.53 | 2.79 | 5.43 | 10.03 |
| Tandem 30-40% | 6 m/s | 2.78 | 3.84 | 6.95 | 10.35 |
| | 10 m/s | 1.92 | 3.13 | 6.16 | 11.42 |

## V. CONCLUSION

Optimization of energy harvesting in wind turbines is of great interest. Here we presented a model for assessment of potential generated energy increment due to using DBD actuators. The phenomenological model which is previously validated by authors [3, 6] was used here. This model was used for detail analysis of the aerodynamic performance of the most significant section. Then simplifying assumptions were used to have a reasonable estimate of the energy increment if the blade, when it was fully equipped with DBD actuators.

First different parts of our simulation tools were validated with other experimental or numerical works. Then flow interactions were simulated in details for seven different single and tandem actuator configurations mounted on the fourth section of the wind turbine blade. It was found that all tandem configurations perform better than single configurations and of course all performed better than the clean airfoil. The physics of the actuator and the boundary flow interactions were analyzed and it was found that the single actuator mounted at 40% the chord length from the leading edge and the tandem actuator mounted at 30-40% of the chord length, had the best performance. The corresponding power increments for these two cases were computed to be, respectively 61 and 95 kW.

Finally, results were applied to all sections to find energy increment due to using these actuators in two different configurations and significant improvements (up-to 10.35% increment for low speed range and 11.42% for high speed range) in energy harvesting at air-speeds of respectively 6 and 10 m/s were observed. This allows to have a higher performance for low wind speeds, which will significantly increase the average output power.